\documentstyle[12pt]{article} \textwidth=17cm \textheight=22.5cm

\def\beq{\begin{equation}} \def\eeq{\end{equation}} \def\bea{\begin{eqnarray}}
\def\eea{\end{eqnarray}}
\def\bq{\begin{quote}} \def\eq{\end{quote}}

\def\gappeq{\mathrel{\rlap {\raise.5ex\hbox{$>$}} {\lower.5ex\hbox{$\sim$}}}}

\def\lappeq{\mathrel{\rlap{\raise.5ex\hbox{$<$}} {\lower.5ex\hbox{$\sim$}}}}

\begin{document} \pagestyle{empty} \begin{flushright} {CERN-TH/98-410 \\ DFPD-98/TH/52} \end{flushright} 
\vspace*{5mm}
\begin{center} {\bf A Simple Grand Unification View of Neutrino Mixing and Fermion Mass Matrices} \\
\vspace*{1cm}  {\bf Guido Altarelli} \\ \vspace{0.3cm} Theoretical Physics Division, CERN \\ CH -
1211 Geneva 23 \\ and
\\Universit\`a di Roma Tre, Rome, Italy \\{\bf
\vspace{0.3cm} Ferruccio Feruglio} \\ \vspace{0.3cm}  
Universit\`a di Padova
\\
and
\\
I.N.F.N., Sezione di Padova, Padua, Italy\vspace{0.3cm} 
\\ \vspace*{2cm}   {\bf Abstract} \\ \end{center}
\vspace*{5mm} \noindent

Assuming three light neutrinos and the see-saw mechanism we present a
semiquantitative model of fermion masses based on (SUSY) SU(5) and abelian
horizontal charges. A good description of the observed pattern of quark
and lepton masses is obtained. For neutrinos we naturally obtain widely
split masses and large atmospheric neutrino mixing as a consequence of
SU(5)-related asymmetric mass matrices for $d$ quarks and charged leptons.

\vspace*{3cm}  \noindent  

\noindent

\begin{flushleft} CERN-TH/98-410 \\ DFPD-98/TH/52 \\ December 1998 \end{flushleft} \vfill\eject 

\setcounter{page}{1} \pagestyle{plain}


Following the experimental results from Superkamiokande \cite{SK} a lot of attention has been devoted to the
problem of a natural explanation of the observed nearly maximal mixing angle for atmospheric
neutrino oscillations. It is possible that also solar neutrino oscillations occur with a large mixing angle 
\cite{solar}. Large mixing angles are somewhat
unexpected because the observed quark mixings are small and the quark, charged lepton and neutrino mass matrices are to
some extent related in Grand Unified Theories. The challenge is to incorporate the new information on neutrino
mixings in a comprehensive picture of fermion masses. In previous papers
\cite{us1,us2} we have given a general discussion of this problem and have proposed a class of solutions
for a natural explanation of maximal mixing within the framework of the see-saw mechanism \cite{ss}. In this article we
review our strategy and present some further examples of realizations of our approach in the context of Grand
Unified Theories (GUT's).

The experimental status of neutrino oscillations is still very preliminary. While the evidence for the existence
of neutrino oscillations from solar and atmospheric neutrino data is rather convincing by now, the values of the
mass squared differences $\Delta m^2$ and mixing angles are not firmly established. For solar neutrinos, for example,
three possible solutions are still possible \cite{solar}. Two are based on the MSW mechanism \cite{MSW}, one with small
(MSW-SA) and one with large mixing angle (MSW-LA), and one in terms of vacuum oscillations (VO) with large mixing angle,
with the MSW-LA solution being now somewhat less favoured than the other two \cite{solar,giu}. For atmospheric neutrinos the preferred
value of
$\Delta m^2$ is affected by large uncertainties and could still sizeably drift in one sense or the other, but
the fact that the mixing angle is large appears established ($\sin^2{2\theta_{atm}}\gappeq 0.8$) \cite{fogli,hall}.
Another issue which is still open is the claim by the LNSD collaboration of an additional signal of neutrino
oscillations in a reactor experiment \cite{LNSD}. This claim was not sofar supported by a second recent experiment,
Karmen \cite{Karmen}, which at face value contradicts the LNSD result, but the issue is far from being closed. Given the
present experimental uncertainties the theorist has to make some assumptions on how the data will finally look like in the
future. Here we tentatively assume that the LNSD evidence will disappear. If so then we only  have two oscillations
frequencies, which can be given in terms of the three known species of light neutrinos without additional
sterile kinds. We then take for granted that the frequency of atmospheric neutrino oscillations will remain well
separated from the solar neutrino frequency, even for the MSW solutions. The present best values are \cite{solar,fogli,hall} 
$(\Delta m^2)_{atm}\sim 2\cdot 10^{-3}~eV^2$ and $(\Delta m^2)_{MSW-SA}\sim 5\cdot 10^{-6}~eV^2$ or $(\Delta
m^2)_{VO}\sim 10^{-10}~eV^2$. We also assume that the electron neutrino does not participate in the atmospheric
oscillations, which (in absence of sterile neutrinos) are interpreted as nearly maximal
$\nu_{\mu}\rightarrow\nu_{\tau}$ oscillations as indicated by the Superkamiokande \cite{SK} and Chooz \cite{Chooz}
data. However the data do not exclude a
non-vanishing $U_{e3}$ element. In most of the Superkamiokande allowed region the bound by Chooz
\cite{Chooz} amounts to  $|U_{e3}|\lappeq 0.2$ but in the region not covered by Chooz $|U_{e3}|$ could
even be somewhat larger \cite{fogli,hall}. If we neglect CP violation phases and adopt a particular set of sign
conventions, the neutrino mixing matrix U is then fixed by the above assumptions in the form explicitly given in
ref. \cite{us2,bar} (see also \cite{Bal}) in
terms of the solar mixing angle (which can be either very small (MSW-SA: $\sin^2{2\theta_{sun}}\sim 5.5\cdot 10^{-3}$) or
nearly maximal (VO:
$\sin^2{2\theta_{sun}}\sim 0.75$).

Neutrino oscillations imply neutrino masses which in turn demand either the existence of right-handed neutrinos
or lepton number violation or both. Given that the neutrino masses are certainly extremely small, it is really
difficult from the theory point of view to avoid the conclusion that lepton number L must be violated. In fact it
is only in terms of lepton number violation that the smallness of neutrino masses can be connected to the very
large scale where L is violated, of order $M_{GUT}$ or even $M_{Pl}\sim 2.4\cdot 10^{18}~{\rm GeV}$. 
If L is not conserved, even in the absence of
$\nu_R$, Majorana masses can be generated for neutrinos by dimension five operators of the form $O_5=L^T_i
\lambda_{ij}L_j\phi\phi/M$ with $\phi$ being the ordinary Higgs doublet, $\lambda$ a matrix in flavour space and
$M$ a large scale of mass. However we consider that the existence of $\nu_R$ is quite plausible because all GUT
groups larger than SU(5) require them. In particular the fact that $\nu_R$ completes the representation 16 of
SO(10): 16=$\bar 5$+10+1, so that all fermions of each family are contained in a single representation of the unifying
group, is too impressive not to be significant. Thus in the following we assume that there are both $\nu_R$ and
lepton number violation. 

With these assumptions the see-saw mechanism is possible and the resulting neutrino mass matrix is of the
form $L^T_im_{\nu ij}L_j$ with $m_{\nu}=m_D^TM^{-1}m_D$ where $m_D$ and $M$ are the neutrino Dirac matrix,
$\bar R m_DL$, and the Majorana matrix, $\bar R M \bar R^T$, respectively. Here we assume that the additional non
renormalisable terms from $O_5$ are comparatively negligible, otherwise they should simply be added. After elimination of
the heavy right-handed fields, at the level of the effective low energy theory, the two types of terms are
equivalent. In particular they have identical transformation properties under a chiral change of basis in flavour
space. The difference is, however, that in the see-saw mechanism, the Dirac matrix $m_D$ is presumably related to
ordinary fermion masses because they are both generated by the Higgs mechanism and both must obey GUT-induced
constraints. Thus if we assume the see-saw mechanism more constraints are implied. In particular we are led to the
natural hypothesis that
$m_D$ has a largely dominant third family eigenvalue in analogy to $m_t$, $m_b$ and $m_{\tau}$ which are by far the
largest masses among $u$ quarks, $d$ quarks and charged leptons. Once we accept that $m_D$ is hierarchical it is very
difficult to imagine that the effective light neutrino matrix, generated by the see-saw mechanism, could have
eigenvalues very close in absolute value. 

Since neutrino oscillations only measure differences of squared masses, the
observed differences $(\Delta m^2)_{atm}=|m^2_3-m^2_2|\gg (\Delta m^2)_{sun}=|m^2_2-m^2_1|$ could correspond to
A) hierarchical eigenvalues $|m_3| \gg |m_{2,1}|$ (that $m_1$ and $m_2$ are close or very different is irrelevant to our
purposes) or to partial or total near degeneracy: B) $|m_1|\sim |m_2| \gg |m_3|$ or C) $|m_1|\sim |m_2| \sim |m_3|$ (the
numbering 1,2,3 corresponds to our definition of the frequencies as in ref. \cite{us2} and in principle may not coincide
with the family index although this will be the case in the models that we favour). The configurations B) and C) imply a
very precise near degeneracy of squared masses. For example, the case C) is the only one that could in principle accommodate
neutrinos as hot dark matter together with solar and atmospheric neutrino oscillations. We think that it is not at all
clear at the moment that a hot dark matter component is really
needed \cite{kra} but this could be a reason in favour of the fully degenerate solution. Then the common mass
should be around 1-3 eV. The solar frequency could be given by a small 1-2 splitting, while the atmospheric
frequency could be given by a still small but much larger 1,2-3 splitting.  A
strong constraint arises in this case from the non observation of neutrinoless double beta decay which requires
that the $ee$ entry of
$m_{\nu}$ must obey
$|(m_{\nu})_{ee}|\leq 0.46~{\rm eV}$ \cite{dbeta}. As observed in ref. \cite{GG}, this bound can only be 
satisfied if bimixing is realized (that is
double maximal mixing, with solar neutrinos explained by the VO or MSW-LA solutions). But we would need a relative
splitting
$|\Delta m/m|\sim
\Delta m^2_{atm}/2m^2\sim 10^{-3}-10^{-4}$ and a much smaller one for solar neutrinos explained by vacuum
oscillations:
$|\Delta m/m|\sim 10^{-10}-10^{-11}$. As mentioned above we consider it unplausible that starting from hierarchical
Dirac matrices we end up via the see-saw mechanism into a nearly perfect degeneracy of squared masses. In
conclusion the assumption of hierarchical Dirac masses and the see-saw mechanism naturally leads to a pattern of
type A with $|m_3| \gg |m_2| \gg |m_1|$. Models with degenerate neutrinos (see, for example, refs. \cite{fri}) could be
natural if the dominant contributions directly arise from non renormalisable operators like $O_5$ which are apriori
unrelated to other fermion masses, but we will not explore this possibility here.

Thus we are led to consider models with large effective light neutrino mass splittings and large mixings. In general
large splittings correspond to small mixings because normally only close-by states are strongly mixed. In a 2 by 2
matrix context the requirement of large splitting and large mixings leads to a condition of vanishing determinant.
For example the matrix
\beq
m\propto 
\left[\matrix{
x^2&x\cr
x&1    } 
\right]
\label{md02}
\eeq
has eigenvalues 0 and $1+x^2$ and for $x$ of 0(1) the mixing is large. Thus in the
limit of neglecting small mass terms of order $m_{1,2}$ the demands of large atmospheric neutrino mixing and
dominance of $m_3$ translate into the condition that the 2 by 2 subdeterminant 23 of the 3 by 3 mixing matrix vanishes.
The problem is to show that this vanishing can be arranged in a natural way without fine tuning. We have discussed
suitable possible mechanisms in our previous paper \cite{us2}. We in particular favour a class of models where, in the
limit of neglecting terms of order $m_{1,2}$ and \underline{in the basis where
charged leptons are diagonal}, the Dirac matrix $m_D$, defined by $\bar R m_D L$, takes the approximate form:
\beq
m_D\propto 
\left[\matrix{
0&0&0\cr
0&0&0\cr
0&x&1    } 
\right]~~~~~. 
\label{md0}
\eeq
This matrix has the property that for a generic Majorana matrix $M$ one finds:
\beq
m_{\nu}=m^T_D M^{-1}m_D\propto 
\left[\matrix{
0&0&0\cr
0&x^2&x\cr
0&x&1    } 
\right]~~~~~. 
\label{mn0}
\eeq
The only condition on $M^{-1}$ is that the 33 entry is non zero. It is important for the following discussion to
observe that $m_D$ given by eq. (\ref{md0}) under a change of basis transforms as $m_D\to V^{\dagger} m_D U$ where V and U
rotate the right and left fields respectively. It is easy to check that in order to make $m_D$ diagonal we need large
left mixings. More precisely $m_D$ is diagonalized by taking V=1 and U given by
\beq
U=  \left[
\matrix{
c&-s&0\cr
sc_{\gamma}&cc_{\gamma}&-s_{\gamma}\cr
ss_{\gamma}&cs_{\gamma}&c_{\gamma}  } 
\right]~~~~~, 
\label{uu}
\eeq
with 
\beq
s_{\gamma}=-x/r~~,~~~~~c_{\gamma}=1/r~~,~~~~~r=\sqrt{1+x^2}~~~~~.
\label{scr}
\eeq
The matrix $U$ is directly the neutrino
mixing matrix. The mixing angle for atmospheric neutrino oscillations is given by:
\beq
\sin^2{2\theta}=4s^2_{\gamma}c^2_{\gamma}=\frac{4 x^2}{(1+x^2)^2}~~~~~.
\label{sin}
\eeq
Thus the bound $\sin^2{2\theta}\gappeq0.8$ translates into  $0.6\lappeq |x|\lappeq 1.6$. It is interesting to
recall that in refs. \cite{ellis,hagi} it was shown that the mixing angle can be amplified by the running from a
large mass scale down to low energy.

We have seen that, in order to explain in a natural way widely split light neutrino masses together with large mixings,
we need an automatic vanishing of the 23 subdeterminant. This in turn is most simply realized by allowing some large
left-handed mixing terms in the Dirac neutrino matrix. By left-handed mixing we mean non diagonal matrix elements that
can only be eliminated by a large rotation of the left-handed fields. Thus the question is how to
reconcile large left-handed mixings in the leptonic sector with the observed near diagonal form of $V_{CKM}$, the quark
mixing matrix. Strictly speaking, since $V_{CKM}=U^{\dagger}_u U_d$, the individual matrices $U_u$ and $U_d$ need not be
near diagonal, but $V_{CKM}$ does, while the analogue for leptons apparently cannot be near diagonal. However nothing
forbids for quarks that, in the basis where $m_u$ is diagonal, the $d$ quark matrix has large non diagonal terms that can
be rotated away by a pure right-handed rotation. We suggest that this is so and that in some way right-handed mixings for
quarks correspond to left-handed mixings for leptons.

In the context of (Susy) SU(5) \cite{ross} there is a very attractive hint of how the present mechanism can be realized. In
the
$\bar 5$ of SU(5) the $d^c$ singlet appears together with the lepton doublet $(\nu,e)$. The $(u,d)$ doublet and $e^c$
belong to the 10 and $\nu^c$ to the 1 and similarly for the other families. As a consequence, in the simplest model with
mass terms arising from only Higgs pentaplets, the Dirac matrix of down quarks is the transpose of the charged lepton
matrix:
$m^d_D=(m^l_D)^T$. Thus, indeed, a large mixing for right-handed down quarks corresponds to a large left-handed mixing
for charged leptons. In the same simplest approximation with  5 or $\bar 5$ Higgs, the up quark mass matrix is
symmetric, so that left and right mixing matrices are equal in this case
\footnote{Up to a diagonal matrix of phases.}. Then small mixings for up quarks and small
left-handed mixings for down quarks are sufficient to guarantee small $V_{CKM}$ mixing angles even
for large $d$ quark right-handed mixings. When the charged lepton matrix is diagonalized the large left-handed mixing
of the charged leptons is transferred to the neutrinos. Note that in SU(5) we can diagonalize the $u$ mass matrix by a
rotation of the fields in the 10, the Majorana matrix $M$ by a rotation of the 1 and the effective light neutrino matrix
$m_\nu$ by a rotation of the $\bar 5$. In this basis the $d$ quark mass matrix fixes $V_{CKM}$ and the charged lepton mass
matrix fixes neutrino mixings. It is well known that a model where the down and the charged lepton mass matrices are exactly
the transpose of one another cannot be exactly true because of the $e/d$ and
$\mu/s$ mass ratios \cite{ross}. It is also known that one remedy to this problem is to add some Higgs component in the 45
representation of SU(5) \cite{jg}. A different solution \cite{eg} will be described later. But the symmetry under
transposition can still be a good guideline if we are only interested in the order of magnitude of the matrix entries and
not in their exact values. Similarly, the Dirac neutrino mass matrix
$m_D$ is the same as the up quark mass matrix in the very crude model where the Higgs pentaplets come from a pure 10
representation of SO(10):
$m_D=m^u_D$. For $m_D$ the dominance of the third family eigenvalue  as well as a near diagonal form could be an order
of magnitude remnant of this broken symmetry. Thus, neglecting small terms, the neutrino Dirac matrix in the basis where
charged leptons are diagonal could be directly obtained in the form of eq. (\ref{md0}).

We give here an explicit example of the mechanism under discussion in the framework of a unified Susy $SU(5)$ theory
with an additional 
$U(1)_F$ flavour symmetry \cite{fro}. This model is to be taken as merely indicative, in that some important problems,
like, for example, the cancellation of chiral anomalies are not tackled here. But we find it impressive that the general
pattern of all what we know on fermion masses and mixings is correctly reproduced at the level of orders of magnitude. 
We regard the present model as a low-energy effective theory valid
at energies close to $M_{GUT}\ll M_{Pl}$. We can think to obtain 
it by integrating out the heavy modes from an
unknown underlying fundamental theory defined at an energy scale 
close to $M_{Pl}$. From this point of view the gauge anomalies
generated by the light supermultiplets listed below 
can be compensated by another set of supermultiplets
with masses above $M_{GUT}$, already eliminated from the 
low-energy theory. In particular, we assume that these 
additional supermultiplets are vector-like with respect to
$SU(5)$ and chiral with respect to $U(1)_F$. Their masses
are then naturally expected to be of the order of the $U(1)_F$
breaking scale, which, in the following discussion, turns
out to be near $M_{Pl}$. We have explicitly checked the 
possibility of canceling the gauge anomalies in this way
but, due to our ignorance about the fundamental theory, 
we do not find particularly instructive to illustrate
the details here. In
this model the known generations of quarks and leptons are contained in  triplets
$\Psi_{10}^a$ and
$\Psi_{\bar 5}^a$, $(a=1,2,3)$ transforming as $10$ and ${\bar 5}$ of $SU(5)$, respectively. Three more $SU(5)$ singlets
$\Psi_1^a$ describe the right-handed neutrinos. We assign to these fields the following $F$-charges:
\bea
\Psi_{10}     & \sim & (3,2,0) \label{c10}\\
\Psi_{\bar 5} & \sim & (3,0,0) \label{c5b}\\
\Psi_1        & \sim & (1,-1,0) \label{c1}
\eea   We start by discussing the Yukawa coupling allowed by $U(1)_F$-neutral  Higgs multiplets $\varphi_5$ and
$\varphi_{\bar5}$ in the $5$ and ${\bar 5}$ $SU(5)$ representations and by a pair $\theta$ and
${\bar\theta}$ of $SU(5)$ singlets with $F=1$ and $F=-1$, respectively. 

In the quark sector we obtain 
\footnote{In eq. (\ref{mquark}) the entries denoted by 1 in $m_D^u$ and $m_D^d$ 
are not necessarily equal. As usual, such a notation allows 
for O(1) deviations.}:
\beq m_D^u=(m_D^u)^T=
\left[
\matrix{
\lambda^6&\lambda^5&\lambda^3\cr
\lambda^5&\lambda^4&\lambda^2\cr
\lambda^3&\lambda^2&1}
\right]v_u~~,~~~~~~~ m_D^d=
\left[
\matrix{
\lambda^6&\lambda^5&\lambda^3\cr
\lambda^3&\lambda^2&1\cr
\lambda^3&\lambda^2&1}
\right]v_d~~,
\label{mquark}
\eeq 
from which we get the order-of-magnitude relations:
\bea m_u:m_c:m_t & = &\lambda^6:\lambda^4:1 \nonumber\\ m_d:m_s:m_b & = &\lambda^6:\lambda^2:1
\eea and 
\beq V_{us}\sim \lambda~,~~~~~ V_{ub}\sim \lambda^3~,~~~~~ V_{cb}\sim \lambda^2~.
\eeq 

Here $v_u\equiv\langle \varphi_5 \rangle$, 
$v_d\equiv\langle \varphi_{\bar 5} \rangle$  and $\lambda$ denotes the ratio between the vacuum expectation value of
${\bar\theta}$ and an ultraviolet cut-off identified with the Planck mass $M_{Pl}$: $\lambda\equiv\langle
{\bar\theta}\rangle/M_{Pl}$. To correctly reproduce the observed quark mixing angles, we  take $\lambda$ of the order
of the Cabibbo angle. For non-negative $F$-charges, the elements of  the quark mixing matrix $V_{CKM}$ depend only on the
charge differences  of the left-handed quark doublet \cite{fro}. Up to a constant shift, this defines the choice in eq.
(\ref{c10}). Equal $F$-charges for $\Psi_{\bar 5}^{2,3}$ (see eq. (\ref{c5b})) are then required to fit
$m_b$ and $m_s$. We will comment on the lightest quark masses later on.

At this level, the mass matrix for the charged leptons  is the transpose of $m_D^d$:
\beq m_D^l=(m_D^d)^T
\eeq and we find:
\beq m_e:m_\mu:m_\tau  = \lambda^6:\lambda^2:1 
\eeq The O(1) off-diagonal entry of $m_D^l$ gives rise to a large left-handed  mixing in the 23 block which corresponds
to a large right-handed mixing in the $d$ mass matrix. In the neutrino sector, the Dirac and Majorana mass matrices are
given by:
\beq m_D=
\left[
\matrix{
\lambda^4&\lambda&\lambda\cr
\lambda^2&\lambda'&\lambda'\cr
\lambda^3&1&1}
\right]v_u~~,~~~~~~~~ M=
\left[
\matrix{
\lambda^2&1&\lambda\cr 1&\lambda'^2&\lambda'\cr
\lambda&\lambda'&1}
\right]{\bar M}~~,
\eeq where $\lambda'\equiv\langle\theta\rangle/M_{Pl}$ and ${\bar M}$  denotes the large mass scale associated to the
right-handed neutrinos: ${\bar M}\gg v_{u,d}$.

After diagonalization of the charged lepton sector and after integrating out the heavy right-handed neutrinos we obtain
the following neutrino mass matrix in the low-energy effective theory:
\beq m_\nu=
\left[
\matrix{
\lambda^6&\lambda^3&\lambda^3\cr
\lambda^3&1&1\cr
\lambda^3&1&1}\right]{v_u^2\over {\bar M}}
\label{mnu}
\eeq where we have taken $\lambda\sim\lambda'$. The O(1) elements in the 23 block are produced by combining the large 
left-handed mixing induced by the charged lepton sector and the large left-handed mixing in $m_D$. A crucial property of
$m_\nu$ is that, as a result of the sea-saw mechanism and of the specific $U(1)_F$ charge assignment, the determinant of
the 23 block is $\underline{automatically}$ of $O(\lambda^2)$ (for this the presence of negative charge values, leading
to the presence of both $\lambda$ and $\lambda'$ is essential \cite{us2}).

It is easy to verify that the eigenvalues of $m_\nu$ satisfy  the relations:
\beq m_1:m_2:m_3  = \lambda^4:\lambda^2:1~~.
\eeq The atmospheric neutrino oscillations require 
$m_3^2\sim 10^{-3}~{\rm eV}^2$. From eq. (\ref{mnu}), taking $v_u\sim 250~{\rm GeV}$, the mass scale ${\bar M}$ of the
heavy Majorana neutrinos turns out to be close to the unification scale, 
${\bar M}\sim 10^{15}~{\rm GeV}$. The squared mass difference between the lightest states is  of $O(\lambda^4)~m_3^2$,
appropriate to the MSW solution  to the solar neutrino problem. Finally, beyond the large mixing in the 23 sector,
corresponding to $s_\gamma\sim c_\gamma$ in eq. (\ref{uu}),
$m_\nu$  provides a mixing angle $s \sim (\lambda/2)$ in the 12 sector, close to the range preferred by the small angle
MSW solution. In general $U_{e3}$ is non-vanishing, of $O(\lambda^3)$.

In general, the charge assignment under 
$U(1)_F$ allows for non-canonical kinetic terms that represent an additional source of mixing. Such terms are allowed by
the underlying flavour symmetry and it would be unnatural to tune them to the canonical form. We have checked that all
the results quoted up to now  remain unchanged after including the effects related to the most general kinetic terms,
via appropriate  rotations and rescaling in the flavour space (see also ref.\cite{lns}). 

Obviously, the order of magnitude description offered by this model is not intended to account for all the details of
fermion masses. Even neglecting the parameters associated with the $CP$ violating observables, some of the relevant
observables are somewhat marginally reproduced.   For instance we obtain $m_u/m_t\sim \lambda^6$  which is perhaps too
large. However we find it remarkable that in such a simple scheme most of the 12 independent  fermion masses and the 6
mixing angles turn out to have the correct order of magnitude.
Notice also that our model prefers large
values of $\tan\beta\equiv v_u/v_d$. This is a consequence
of the equality $F(\Psi_{10}^3)=F(\Psi_{\bar 5}^3)$ (see eqs. (\ref{c10}) and (\ref{c5b})).
In this case the Yukawa couplings of top and bottom quarks 
are expected to be of the same order of magnitude, while the large
$m_t/m_b$ ratio is attributed to $v_u \gg  v_d$ (there may be factors
O(1) modifying these considerations, of course). We recall here that
in supersymmetric grand unified models large values of $\tan\beta$ 
are one possible solution to the problem of reconciling the boundary 
condition $m_b=m_\tau$ at the 
GUT scale with the low-energy data \cite{largetb}. 
Alternatively, to keep $\tan\beta$ small, one could 
suppress $m_b/m_t$ by adopting different $F$-charges for the 
$\Psi_{\bar 5}^3$ and $\Psi_{10}^3$.

Additional contributions to flavour changing processes and to
$CP$ violating observables are generally expected in a supersymmetric
grand unified theory.
However, a reliable estimate of the corresponding effects
would require a much more detailed definition of the theory
than attempted here. Crucial ingredients such as the mechanism
of supersymmetry breaking and its transmission to the observable
sector have been ignored in the present note. We are implicitly
assuming that the omission of this aspect of the
flavour problem does not substantially alter our discussion. 

A common problem of all $SU(5)$ unified theories based on a minimal higgs structure is represented by the relation
$m_D^l=(m_D^d)^T$ that, while leading to the successful $m_b=m_\tau$ boundary condition at the GUT scale, provides the
wrong prediction
$m_d/m_s=m_e/m_\mu$ (which, however, is an acceptable order of magnitude equality). We can easily overcome this problem
and improve the picture \cite{eg} by introducing an additional supermultiplet
${\bar\theta}_{24}$ transforming in the adjoint representation of $SU(5)$ and  possessing a negative $U(1)_F$ charge,
$-n~~(n>0)$.  Under these conditions, a positive
$F$-charge $f$ carried by the matrix elements 
$\Psi_{10}^a \Psi_{\bar 5}^b$ can be compensated  in several different ways by monomials of the kind
$({\bar\theta})^p({\bar\theta}_{24})^q$, with
$p+n q=f$. Each of these possibilities represents an independent contribution to the down quark and charged lepton mass
matrices, occurring with an unknown coefficient of O(1). Moreover the product $({\bar\theta}_{24})^q \varphi_{\bar 5}$
contains both the ${\bar 5}$ and the $\overline{45}$ $SU(5)$ representations,  allowing for a differentiation between
the down quarks and the charged leptons. The only, welcome, exceptions are  given by the O(1) entries that do not
require any compensation and, at the leading order, remain the same for charged leptons and  down quarks. This preserves
the good $m_b=m_\tau$ prediction. Since a perturbation of O(1) in the subleading matrix elements  is sufficient to cure
the bad $m_d/m_s=m_e/m_\mu$ relation, we can safely assume that 
$\langle{\bar\theta}_{24}\rangle/M_{Pl}\sim\lambda^n$, to preserve the correct order-of-magnitude predictions in the
remaining sectors. 

We have not dealt here with the problem of recovering the correct 
vacuum structure by minimizing the effective potential of the theory.
It may be noticed that the presence of two multiplets $\theta$ and
${\bar \theta}$ with opposite $F$ charges could hardly be reconciled,
without adding extra structure to the model,
with a large common VEV for these fields, due to possible analytic
terms of the kind $(\theta {\bar \theta})^n$ in the superpotential
\footnote{We thank N. Irges for bringing our attention on this point.}.
We find therefore instructive to explore the consequences of allowing
only the negatively charged ${\bar \theta}$ field in the theory.

It can be immediately recognized that, while the quark mass matrices
of eqs. (\ref{mquark}) are unchanged, in the neutrino sector the Dirac 
and Majorana matrices get modified into:
\beq
m_D=
\left[
\matrix{
\lambda^4&\lambda&\lambda\cr
\lambda^2&0&0\cr
\lambda^3&1&1}
\right]v_u~~,~~~~~~~~
M=
\left[
\matrix{
\lambda^2&1&\lambda\cr
1&0&0\cr
\lambda&0&1}
\right]{\bar M}~~.
\eeq
The zeros are due to the analytic property of the superpotential
that makes impossible to form the corresponding $F$ invariant
by using ${\bar \theta}$ alone.
These zeros should not be taken literally, as they will be eventually  
filled by small terms coming, for instance, from the diagonalization
of the charged lepton mass matrix and from the transformation
that put the kinetic terms into canonical form. It is however
interesting to work out, in first approximation, the case 
of exactly zero entries in $m_D$ and $M$, when forbidden by $F$.

The neutrino mass matrix obtained via see-saw from $m_D$ and $M$
has the same pattern as the one displayed in eq. (\ref{mnu}).
A closer inspection reveals that the determinant of the 23 block
is identically zero, independently from $\lambda$.
This leads to the following pattern of masses:
\beq
m_1:m_2:m_3  = \lambda^3:\lambda^3:1~~,~~~~~m_1^2-m_2^2 = 
{\rm O}(\lambda^9)~m_3^2~~.
\eeq
Moreover the mixing in the 12 sector is almost maximal:
\beq
{s\over c}={\pi\over 4}+{\rm O}(\lambda^3)~~.
\eeq
For $\lambda\sim 0.2$, both the squared mass difference $(m_1^2-m_2^2)/m_3^2$ 
and $\sin^2 2\theta_{sun}$ are remarkably close to the values 
required by the vacuum oscillation solution to the solar neutrino
problem. We have also checked that this property is reasonably stable
against the perturbations induced by small terms (of order $\lambda^5$)
replacing the zeros,
coming from the diagonalization of the charged lepton sector 
and by the transformations that render the kinetic terms canonical.
We find quite interesting that also the just-so solution, requiring 
an intriguingly small mass difference and a bimaximal mixing,
can be reproduced, at least at the level of order of magnitudes,
in the context of a "minimal" model of flavour compatible with
supersymmetric SU(5). In this case the role played by supersymmetry 
is essential, a non-supersymmetric model with ${\bar \theta}$ alone 
not being distinguishable from the version with both $\theta$
and ${\bar \theta}$, as far as low-energy flavour properties are concerned.

Finally, let us compare our model with other recent proposals \cite{Large}. Textures for the effective neutrino mass matrix similar
to $m_\nu$ in eq. (\ref{mnu}) were derived in refs. \cite{ram,buch}, also in the context of an $SU(5)$ unified theory
with a $U(1)$ flavour symmetry. In these works, however, the O(1) entries of
$m_\nu$ are uncorrelated, due to the particular choice of U(1) charges. The diagonalization of such matrix, for generic
O(1) coefficients, leads to only one  light eigenvalue and to two heavy eigenvalues, of O(1), in units of 
$v_u^2/{\bar M}$. Then the required pattern $m_3\gg m_2\sim m_1$ has to be fixed by hand. On the contrary, in
 our model the desired pattern is automatic, since, as emphasized above, the determinant of the 23 block in $m_\nu$
is vanishing at the leading  order. Other models in terms of U(1) horizontal charges have been proposed in refs.
\cite{abelian,froggart,ellis}. Clearly a large mixing for the light neutrinos can
be provided in part by the diagonalization of the charged lepton sector. As we have seen, in
$SU(5)$, the left-handed mixing  carried by charged leptons is expected to be, at least in first approximation, directly
linked to the right-handed mixing for the $d$ quarks and, as such, perfectly compatible with the available data. This
possibility was remarked, for instance, in refs. \cite{bere0,alb,bere,hagi} where the implementation was in terms of
asymmetric textures, of the Branco et al. type \cite{bra}, used as a general parametrization of the  existing data
consistent with the constraints imposed by the unification program. On the other hand, our model aims to a dynamical
explanation of the flavour properties, although in a simplified setting 
\footnote{Also ref. \cite{bere} suggests an horizontal $U(2)$ symmetry to justify the assumed textures.}.

Our conclusion is that if we start from three light neutrinos and the see-saw mechanism then the most natural
interpretation of the present data on neutrino oscillations is in terms of hierarchical light neutrino masses and
asymmetric mass matrices (at least for $d$ quarks and charged leptons). As well known, asymmetric matrices allow to
reproduce the experimental value of $|V_{cb}|$ better than in the symmetric Fritzsch texture \cite{frit}. While SO(10),
perhaps realized in some form at
$M_{Pl}$, appears as a good classification group, with each family perfectly accommodated in a 16 representation,
the description at $M_{GUT}$ is more accurately formulated in terms of $SU(5)$. In this framework it is natural to
obtain large splittings and large mixing angles for light neutrinos.   

\noindent
{\bf Acknowledgements}

\noindent
One of us (F.F.) would like to thank Anna Rossi and Fabio Zwirner for useful discussions. We also thank Nikolaos Irges for an interesting observation.

\vfill
\end{document}